\documentstyle[prl,aps,twocolumn,epsf]{revtex}
\begin{document}
\title{\large Sign reversals of the Quantum Hall Effect in quasi-1D
conductors} \author{D. Zanchi and G. Montambaux}
\address{Laboratoire de Physique des Solides,  associ\'e au 
CNRS \\ Universit\'{e} Paris--Sud \\ 91405 Orsay, France}
\twocolumn[
\date{\today}
\maketitle
\widetext
 \begin{center}
 \begin{abstract}
\parbox{14cm}{The sign reversals of the Quantum Hall Effect observed in
quasi-one-dimensional conductors of the Bechgaard salts family are explained
within the framework of the quantized nesting model. The sequence of
reversals is driven by slight modifications of the geometry of the Fermi
surface. It is explained why only even phases can have sign reversals and
why negative phases are less stable than positive ones.
} \end{abstract}
\end{center}
]
\narrowtext
During last years, remarkable properties under magnetic
field have been discovered in a class of quasi-one-dimensional
(quasi-1D) conductors\cite{Review,Montambaux91,Model1,Model2}.  Organic
conductors of the  Bechgaard  salts   family,
 $(TMTSF)_2X$ where $TMTSF$ $=$ tetramethyselenafulvalene, are
strongly  anisotropic systems, with a typical hierarchy of transfer
integrals: $t_a=3000K, t_b= 300K, t_c=10K$. In three
members of this family  ($X = ClO_4, PF_6, ReO_4$), the metallic phase is
 destroyed by a moderate magnetic field $H$ applied along
the $\bf c^*$ direction, perpendicular to the most
 conducting planes $(\bf a,\bf b)$ and a field induced
 phase appears which consists in a series of Spin Density Wave (FISDW)
 subphases, separated by first order transitions.
This field induced cascade of quantized phases results from an interplay
between the nesting properties of the Fermi surface (FS) and the
quantization
of electronic orbits in the field: the wave vector of the SDW adjusts
itself with the field so that unpaired carriers in the subphases always
fill an integer number of Landau levels. As a result, the
number of carriers in each subphase is quantized and so is the Hall
conductivity: $\sigma_{xy}= 2 N e^2/\hbar$ (a factor
 $2$ accounts for spin degeneracy)\cite{Poilblanc86,Yakovenko91}.
The apparition of these phases results from a new structure of the metallic
phase in a field: because of the Lorentz force, the electronic
motion becomes periodic and confined along the direction of the
chains of high conductivity ($\bf a$ direction). As a result of this
effective reduction of dimensionality, the metallic phase becomes
unstable\cite{Model1}. The periodic motion of  the electrons
in real space is  characterized by the
wave vector $G= e H b/\hbar$, $b$ being the interchain distance.
Consequently,
the static spin susceptibility $\chi_0({\bf Q})$, instead of having one
logarithmic divergence
at  $2k_F$, exhibits a series of divergences at quantized values of the
longitudinal component of the wave
vector $Q^n_\parallel =
2 k_F + n G$\cite{Montambaux91,Model2,Montambaux85b}.
 The largest divergence signals the appearance of a SDW
phase with quantized  vector $Q_\parallel = 2 k_F + N G$.
These ideas have been formalized in the so-called Quantized Nesting
Model (QNM) which describes most of the features of the
phase diagram in a magnetic field, in particular the observed Hall
plateaux\cite{Montambaux91,Model2}.

However, one of the most puzzling  unexplained experimental results is
certainly the possibility of a reversal of
the Hall effect  when the field varies: although most of the phases
exhibit
the same sign of the Hall voltage (by convention we will refer to these
plateaus as the {\it positive} ones\cite{Remark}),
 it has been discovered by Ribault that  {\it negative}
plateaus may appear in $(TMTSF)_2ClO_4$  under certain conditions of
cooling rate\cite{Ribault85}.
Such negative plateaus  have been reproduced and also
found in $(TMTSF)_2PF_6$, where their existence crucially depends on the
pressure\cite{Piveteau86,Cooper89,Hannahs89}.
 One of the puzzling aspects of the negative plateaus is that most often
they resemble a dip rather than a plateau and they seem less stable than
 the positive ones.

Quite recently,
  by a conditioning procedure in which current pulses depin the FISDW
from lattice defects and tend to reduce hysteresis,
 it has been shown  unambiguously that
there exists at least one phase characterized by a well-formed negative
plateau with a quantized
value of $\sigma_{xy}$ ($N=-2$)\cite{Balicas95}. In this experiment, the
sequence
of  plateaus  obtained by decreasing the field can be clearly identified
with the quantum numbers $N=1,2,-2,3,4,5,6,7$.
 Although there is only one negative plateau in this experiment, an older
work have shown  a sequence of phases which could be labeled
by $N=1,2,-2,4,-4,5,6$\cite{Cooper89}. Note that the negative plateaus
are  labeled by {\it even} numbers only. In others salts, $ClO_4$ and
$ReO_4$,
there are also several negative features but it is more difficult to ascribe
them a well defined quantum number\cite{Ribault85,Kang91} (
$ReO_4$: $1,2,-2,?$). Moreover, in these two materials, the
situation is complicated by the anion ordering which certainly affects the
apparition of subphases\cite{Osada92}. In $ClO_4$, the existence of
negative phases is also very dependent on pressure\cite{Kang93}.

 In this letter, we show  that these sign reversals  can be described as an
equilibrium solution of the Nesting Model already used to describe the
FISDW when there is no sign reversal.
 It is essential to notice that at least the phase
$N=-2$ extends up to the
metallic state\cite{Balicas95}. This cannot be explained by a theory
based on multiple order
parameter states which would appear only at low temperature\cite{Machida94}.
 On the contrary, a description of the metallic
 phase instability {\it should}  explain the existence of such phases with
an appropriate form of the FS.

In order to describe such phases, two "strategies" are {\it a priori}
possible. One is to use the dispersion
 relation deduced from band structure calculations for real salts with the
triclinic symmetry\cite{Grant83}. We do not think that this
could be the
best method because the phase diagram is sensitive to extremely small energy
scales which may not be included in the band structure
calculations.   The  other method is to model a dispersion
relation with
 the essential ingredients to describe the observed features. This is
the method that we have followed.
We found indeed that an appropriate and very slight modification of the
FS  can lead to the
sequence of plateaus $1,2,-2,3,4,-4,5$. Our result strengthens the
validity  of the QNM used to describe
the phase  diagram of Bechgaard salts in a magnetic field.

The relevant hamiltonian for the metallic phase writes: 
\begin{equation}
{\cal H}=v_F(|k_x|-k_F)+t_\perp(k_y b)
\end{equation}
$t_\perp(k_y b)$ is a periodic function which
 describes a warped FS. It satisfies
the properties $ t_\perp(p+ 2 \pi) = t_\perp(p)$ and $t_\perp(-p) =
t_\perp(p)$. It can be expanded in a Fourier series as:
\begin{equation}   \label{dispersion}
t_\perp(p)=-2t_b\cos p-2t_b^{\prime }\cos 2p-2t_3 \cos 3p - 2 t_4 \cos 4p
\end{equation}
Although  essential to explain the existence of a threshold field
for the cascade of FISDW, the coupling in the third direction is omitted
since it is known that
it does not play an important role in the sequence of
subphases\cite{Montambaux91,Montambaux85a}.
As usual, the dispersion relation is linearized along the direction $x$ of
highest conductivity.
The first harmonics ($t_b$) of the dispersion along the transverse direction
$y$
describes the warping of the FS with a perfect nesting at wave vector $(2
k_F, \pi/b)$. The second harmonics ($t'_b$) induces a deviation
from perfect nesting which leaves a small number of carriers quantized
into Landau bands. Its amplitude fixes the period of the
cascade\cite{Montambaux91,Model2}.
Typically in Bechgaard salts, $t'_b \simeq 10
K$\cite{Montambaux91,Montambaux85a}.
This term may have two origins. One
is  the linearization
of the dispersion relation along the $x$ direction\cite{Yamaji82}.
 Other contributions may result directly from next nearest neighbor
coupling\cite{Yamaji86}.
The "standard"  QNM, which includes only these two
harmonics,
explains the existence of a   cascade of FISDW phases with
$\sigma _{xy}= 2 N e^2/h$, with a given sign of $N$, i.e. no reversal of
the Hall sign.

It is easy to understand qualitatively why the standard model cannot
lead to sign reversals. When the field
varies, the nesting
vector oscillates around its zero field value, which connects the inflexion
points of the FS\cite{Montambaux85a}. The sign of the carriers
is thus given by
the position of the inflexion point. A simple geometric analysis shows that
$sign(N)=sign(Q_\parallel - 2 k_F) =  sign(t'_b)$.
However, the fact that the sign of the Hall sequence is fixed is
certainly not a general
feature. The detailed structure of $\chi _0$ has to depend on the fine
geometry of the FS\cite{Montambaux85a}. If two regions of the FS
exhibit almost equally good nesting properties, one can imagine that the SDW
vector will oscillate from one to another one at
$Q_\parallel > 2 k_F$ and the other at $Q_\parallel < 2 k_F$.

We show here that this can be indeed the case. A very slight
modification of the FS induced by a third ($t_3$) and a fourth
($t_4$) harmonics in the transverse
direction is enough to explain the existence of new phases with a change in
the sign of the Hall effect.   We explain now why these two terms are
both necessary to describe the sign reversals.

We have calculated $\chi_0$ with the general
dispersion relation (\ref{dispersion}) and searched for the
absolute maximum which signals the appearance of each subphase.
It is given by \begin{equation}
\chi_0({\bf Q}) = \sum_n I_n^2(Q_\perp) \chi_0^{1D}(Q_\parallel - n G)
\end{equation}
This forms exhibits the structure
of $\chi_0$ as the sum of one-dimensional
 contributions  $\chi_0^{1D}$ shifted by the magnetic
field wave vector $G= e H b / \hbar$\cite{Montambaux86}.
The $I_n$ depend on the zero field dispersion relation\cite{Montambaux86}.
\begin{equation} \label{In}
I_n(Q_\perp) = \langle
e^{i[T_\perp(p+Q_\perp/2)+T_\perp(p-Q_\perp/2)+n
p]} \rangle \end{equation}
where $T_\perp(p)=(2/\hbar \omega_c) \int_0^p  t_\perp(p') dp'$ and
 $\langle ... \rangle$ is the average over $p$. $\omega_c = e v_F b H/2$
is the cyclotron frequency of the open periodic motion in the metallic phase
and $\hbar \omega_c$ is the separation between Landau bands in the FISDW
phases. $\chi_0$ has logarithmic divergences
at quantized values of the wave vector  and the largest divergence
signal the formation of a FISDW at the corresponding wave vector
 ${\bf Q}_N = (2 k_F + N eHb/\hbar,Q_\perp)$.
When the field is varied, peaks with different $N$ become in turn the
absolute maximum.

The standard model ($t_3=t_4=0$) leads to $\chi_0({\bf Q}_N) > \chi_0({\bf
Q}_{-N})$
for all magnetic fields, so that the sign reversal could not be explained in
this framework, figs. 1 a,b.
 Note  that there are several maxima
for each value of the quantum number and that
there is a local maximum, noted ${\bf Q}^0$, on the $Q_\perp = \pi / b$
line, but  {\it for even phases only} . This is
because $I_{2M+1}(\pi / b) =0$, a property which can be
checked directly
from eq. (\ref{In}). This property has a simple semiclassical qualitative
explanation: when  $Q_\perp = \pi / b$, the pocket of unnested carriers
is  splitted into {\it two} pockets of equal size. Quantization in each of
these two pockets implies an {\it even} quantization of the total number of
unnested carriers. For the standard
model, the absolute maximum, noted $\bf Q^*$, always lies {\it outside} the
$Q_\perp = \pi
/ b$ line\cite{Montambaux85b,Montambaux85a}. This is a reminiscence of the
position of the zero field best nesting vector\cite{Montambaux85a}.
\begin{figure}
\centerline{
\epsfxsize 6cm
\epsffile{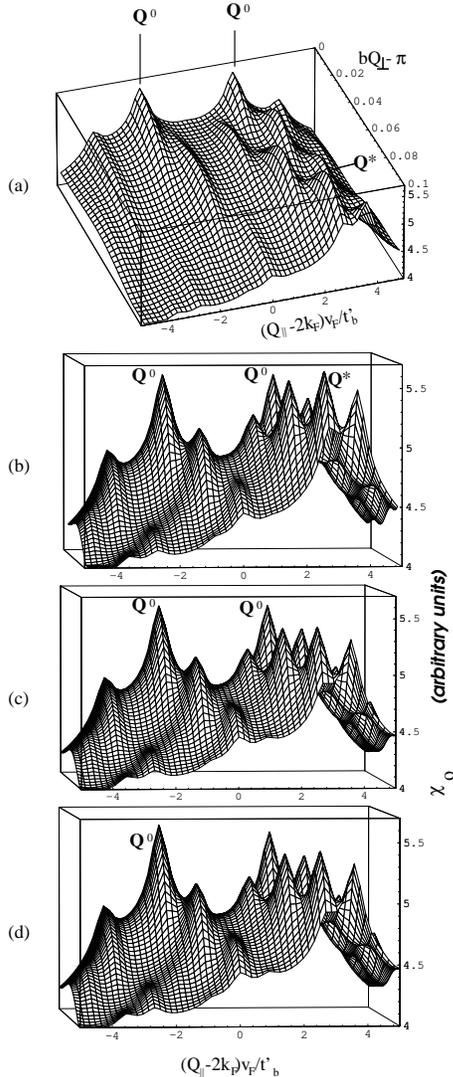}}
\caption{a) $\chi_0({\bf Q})$ in a finite magnetic field for the standard
model, $t'_b=10K,t_3=t_4=0$. The best nesting vector is ${\bf Q}^*$. ${\bf
Q}^0$
is a degenerate secondary maximum. b) Same parameters. c) A finite $t_3=
10K$
alters the best nesting and  ${\bf Q}^0$ is now the degenerate best nesting
vector. d)  A finite $t_4= 0.2K$ lifts the degeneracy,
 leading to a negative quantum number.
}
\label{fig1}
\end{figure}
The effect of a third harmonics $t_3$ in the  dispersion relation is to
deteriorate the nesting at the inflexion point. This is seen on fig.2b
for $H=0$ and on fig.1c for $H \neq 0$. To make the effect more visible on
the figures, we have chosen a large  $t_3 = t'_b$ but we have found
that
a smaller value of $t_3 \approx 0.2 t'_b$ is sufficient. As a result, when
the
field is applied, the odd absolute maxima  have  still  $Q_\parallel >2 k_F$
    but the even maxima can be of two different natures depending on the
 field:  either they stand on the $Q_\perp = \pi / b$
line ($Q^0$ on fig.1) or towards the zero field best nesting vector
  ($Q^*$ on fig. 1). When they lie on the $Q_\perp = \pi / b$ line,
these  maxima are degenerate: $\chi_0({\bf Q}^0_{2M})
= \chi_0({\bf Q}^0_{-2M})$. This degeneracy had already been noticed in the
past\cite{Machida94,Montambaux88}.   It would lead to a phase diagram
where $-2$ and $2$ are degenerate, which is not the case.
\begin{figure}
\centerline{
\epsfxsize 6cm
\epsffile{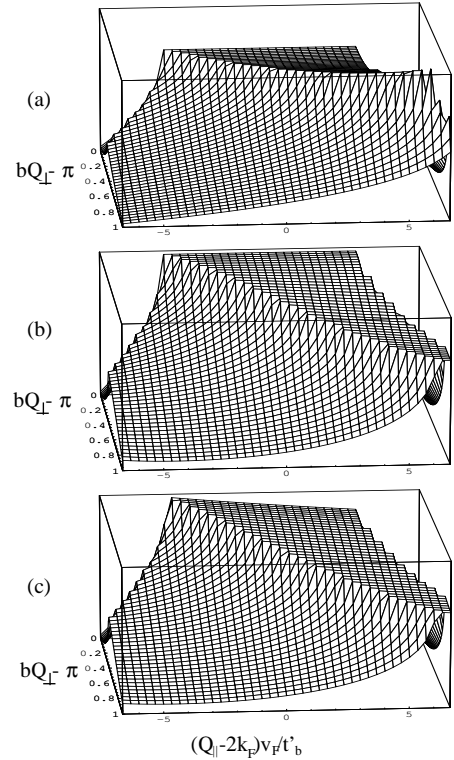}}
\caption{a) $\chi_0({\bf Q})$ in zero field for the
standard model. There
is a  maximum corresponding to the inflexion point of the FS
with $Q_\parallel > 2k_F$.
b) When $t_3 \neq 0$,  the maximum is moved along  the degenerate $Q_\perp
= \pi /b$ line. c) When  $t_4 \neq 0$, the maximum has $Q_\parallel <
2k_F$. To increase the effect on the figures, we have chosen large values
of the paremeters $t'_b=60K$, $t_3=20K$, $t_4=2K$.} \label{fig2}
\end{figure}
 We have found
that this degeneracy can  be removed by the addition of a fourth harmonics
of amplitude $t_4$. On the  $Q_\perp = \pi / b$ line, the $I_N$ are given
by: \begin{equation}
I_{N}({\pi \over b}) = \langle \exp [ {4 i  \over \hbar \omega_c} (t'_b \sin
2p - {t_4 \over 2} \sin 4p) +  i N p] \rangle
\end{equation}
By changing $N$ into $-N$ and $p$ into $p + \pi /2$ ,
one has:
\begin{equation}
I_{-N}({\pi \over  b}) = (-1)^{{N\over2 }}\langle \exp [ {4 i  \over \hbar
\omega_c} (t'_b \sin 2p + {t_4 \over 2} \sin 4p) +  i N p] \rangle
\end{equation}
One immediately sees that if $N$ is odd, $I_N=0$ as stated above. If $N$ is
even, the degeneracy is broken by a non-zero $t_4$ (figs.2c,1d). When
$sign(t_4) = sign(t'_b)$, $I_{-N}^2 > I_{N}^2$, so that $\chi_0({\bf
Q}_{-N} ) > \chi_0({\bf Q}_{N} )$ and a phase with  {\it negative even}
$N$ is favored.
We have found the best nesting vector  as a function of the magnetic field.
From the obtained sequence of quantized nesting vectors vs. field, we have
obtained
the  evolution of the Hall voltage shown on fig.3. It is very similar
 to the experimental one\cite{Cooper89}.
At low $T$, a sequence of fine quantized
 structures may be
resolved and an alternance of many subphases
with positive and negative quantum numbers can appear.
We think that this explains the series of oscillations in Hall
effect observed near the threshold field a few years ago\cite{Piveteau86}.
Depending on the value of the critical temperature $T_c(H)$, different
sequences can be obtained. For higher $T_c(H)/\hbar \omega_c$, the low field
oscillations can disappear and only plateaux with small $N$
remain well-defined.
\begin{figure} \centerline{ \epsfxsize 6cm
\epsffile{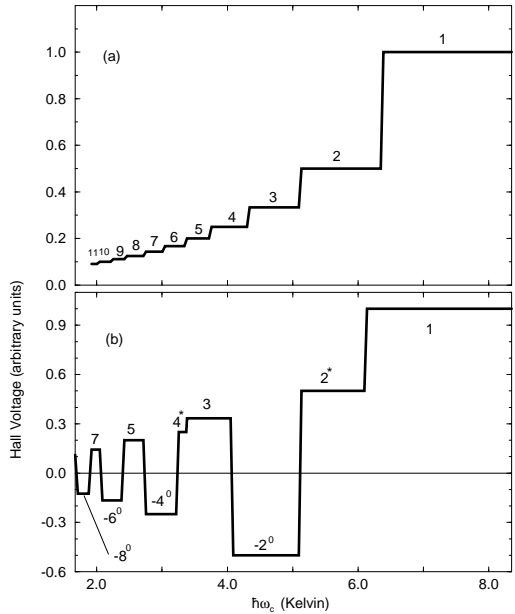}}
\caption{ Hall voltage versus field for a)  $t'_b = 10K$,  $t_3 = t_4=0$
  b) $t_3 = 7K$,
  $t_4 =.025K$, obtained from $\chi_0(\bf Q)$ at $T=0.5K$.
}
\label{fig3}
\end{figure}
 We believe that the
experimentally observed high sensitivity of the sequence of sign reversal
to external pressure is due to the sensitivity of the parameters of the
dispersion relation to pressure. We emphasize that although
 the metallic electron gas  has a metallic behavior with a large Fermi energy,
 of the order of the $eV$, the cascade of the FISDW's
 is driven by extremely small energy scales, of the order of a few Kelvins
or even less.  The orders of magnitude of the two additional harmonics
($t_3$ and $t_4$) are not incompatible
with the estimations of  a refined microscopic  model\cite{Yamaji86}

Finally it is worth noticing that positive and negative even phases are
almost
degenerate because the energy scale $t_4$ is certainly very small, of the
order of $1K$ or less. The relative energy difference between these two
phases is very small, of order $(t_4/\omega_c)^2$. This explains why the
negative phases are always very
sensitive to external parameters like pressure or probably anion ordering.

In conclusion, we have explained the ten-years-old puzzle of the
observed
sign reversals of the Quantum Hall effect in the cascade of FISDW phases of
Quasi-1D organic conductors. They can be described
 with a slight modification of the dispersion relation of the
metallic phase.
We have been able to reproduce the observed sequnce of "negative" phases
with an even quantum number, to understand why they are very sensitive to
pressure and why it is more difficult to measure a well defined plateau.
  Our result shows that the electronic properties of the
Bechgaard salts are extremely sensitive to very small changes in the
geometry of the Fermi surface and that the QNM and its variations still
continues to describe very well the observed phase diagram of these salts in
a magnetic field.

Acknowledgments:   We thank  L. Balicas, A. Bjeli\v{s}, M. H\'eritier, D.
J\'erome,  G. Kriza and  M.
Ribault for very useful discussions.


\begin{thebibliography}{99}
\bibitem{Remark}In this letter, we emphasize on the sign reversal of the
 Hall effect but not on the "absolute" sign of the Hall
 effect which is not clearly defined from the literature. In the nesting
 model, this sign depends on the sign of $t'_b$.
\end{thebibliography}
\end{document}